# Distributed Systems in Fintech

**Anurag Mashruwala**
anuragm@bu.edu

**Abstract.** The emergence of distributed systems has revolutionized the financial technology (Fintech) landscape, offering unprecedented opportunities for enhancing security, scalability, and efficiency in financial operations. This paper explores the role of distributed systems in Fintech, analyzing their architecture, benefits, challenges, and applications. It examines key distributed technologies such as blockchain, decentralized finance (DeFi), and distributed ledger technology (DLT), and their impact on various aspects of the financial industry, and future directions for distributed systems in Fintech.

# 1 Introduction

## 1.1 Overview of Fintech

Fintech, short for Financial Technology, refers to the use of innovative technology to deliver financial services and products in more efficient, accessible, and cost-effective ways. It encompasses a broad range of applications, including but not limited to online banking, mobile payment systems, peer-to-peer lending platforms, cryptocurrency, robo-advisors, and blockchain technology.

It represents the intersection of finance and technology, leveraging advancements such as artificial intelligence, big data analytics, cloud computing, and mobile technology to transform traditional financial services and create new opportunities for consumers, businesses, and financial institutions.

Understanding Fintech requires consideration within the framework of the digital finance cube. This cube consists of three dimensions: finance functions, the technologies utilized for these functions, and the institutions engaged in digitizing finance functions [1].

Hence, within this framework, Fintech can be characterized as the endeavors focused on digitalizing finance functions (such as financing, investments, money management, payments, insurance, and financial advisory services) through the utilization of finance technologies. These technologies are either adopted or being developed by specific organizations within both the finance and IT sectors [2].

## 1.2 Introduction to distributed systems and their relevance in Fintech

A distributed system is a network of interconnected computers or nodes that work together to achieve a common goal. In a distributed system, tasks are divided among multiple nodes, allowing them to work simultaneously and independently. These nodes communicate with each other through a network, sharing resources and data to accomplish tasks efficiently. Unlike centralized systems, where all processing and data storage occur on a single machine, distributed systems distribute these functions across multiple nodes, providing benefits such as fault tolerance, scalability, and improved performance.

Distributed systems play a pivotal role in the realm of Fintech, serving as the backbone for numerous innovations and applications within the financial technology sector. These systems, characterized by their decentralized architecture and collaborative nature, offer a range of capabilities that are particularly well-suited to the needs and challenges of modern finance. By distributing tasks and data across multiple nodes interconnected through networks, distributed systems ensure high availability, scalability, and fault tolerance, essential qualities for mission-critical financial services. In the context of Fintech, where uninterrupted availability and rapid scalability are paramount, distributed systems enable platforms to handle increasing workloads, accommodate growing user bases, and withstand potential hardware failures or network disruptions. Moreover, distributed systems facilitate secure and efficient data management and processing, vital for maintaining data consistency, integrity, and confidentiality in financial transactions and operations. With the rise of digital payments, blockchain technology, peer-to-peer lending platforms, and algorithmic trading systems, distributed systems provide the infrastructure and capabilities necessary to support these innovations, offering secure, scalable, and resilient solutions for the evolving needs of the financial industry. Distributed systems are thus integral to the success and advancement of Fintech, driving innovation, efficiency, and reliability in financial services and applications.

## 2. Distributed Systems

### 2.1 What are distributed systems?

There are several definitions of distributed systems found in the literature. Some of the definitions are:

'A distributed system is a collection of autonomous computing elements that appears to its users as a single coherent system' [3, p. 968]

'A distributed system is a collection of processes working together to accomplish some task.' [4, p. 122]

'…distributed systems are systems in which cooperating entities do not share the same physical space and/or do not have a common time reference.' [5, p. 155]

While there doesn't appear to be a rigid definition of distributed systems we can characterize distributed systems as systems having numerous independent computational units, such as computers or nodes that co-ordinate their actions to achieve a common goal. These systems typically have a networked architecture and exchange messages to function. A distributed system is shown in Figure 1 where the nodes are autonomous computing units.

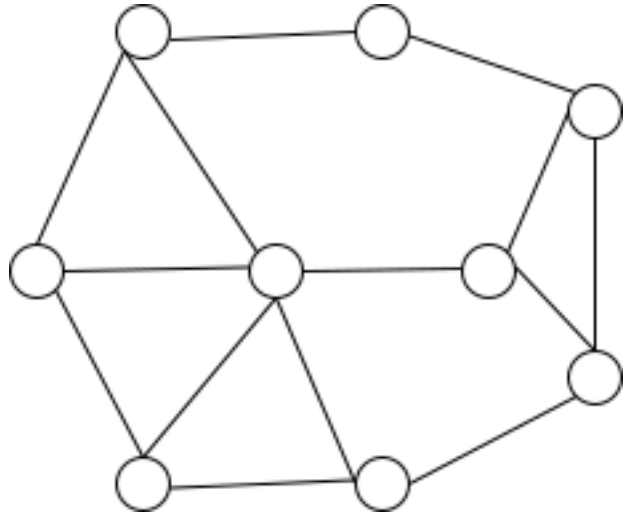

Figure 1

## 2.2 Explanation of distributed systems architecture

As explained above, a distributed system is a network of autonomous computers or nodes, each with its own processor and memory, that communicate with each other through a network. In a distributed system, these nodes work together to achieve a common goal, such as processing data, running applications, or providing services to users. Let us look at common architectures that are employed by distributed system architects:

**Client-Server Architecture**: This is one of the most basic architectures where clients make requests to servers, which then process these requests and send back responses. It's widely used in many applications, including web servers where clients (web browsers) request web pages from servers.

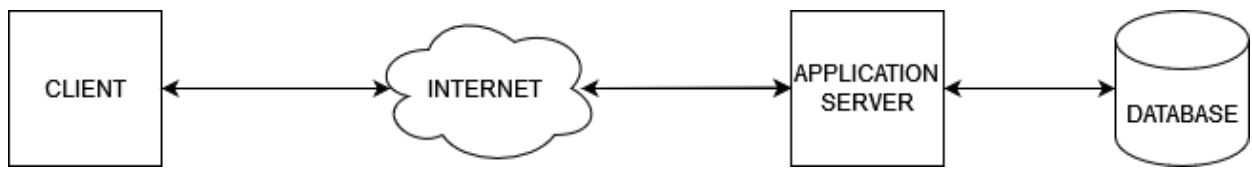

Figure 2

**Peer-to-Peer (P2P) Architecture**: In this architecture, all nodes can act as both clients and servers. Each node can initiate requests and respond to requests from other nodes directly, without the need for a central server. P2P networks are often used for file sharing, communication, and distributed computing.

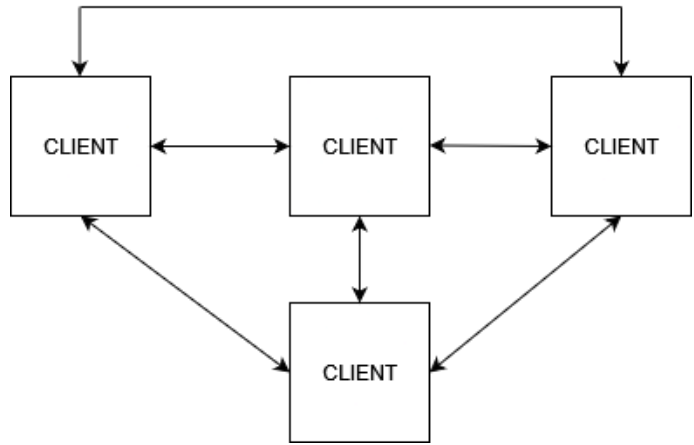

Figure 3

**Microservices Architecture**: This architecture decomposes an application into a set of small, independently deployable services, each running in its own process and communicating with other services through lightweight mechanisms like HTTP or messaging queues. Microservices allow for better scalability, flexibility, and resilience.

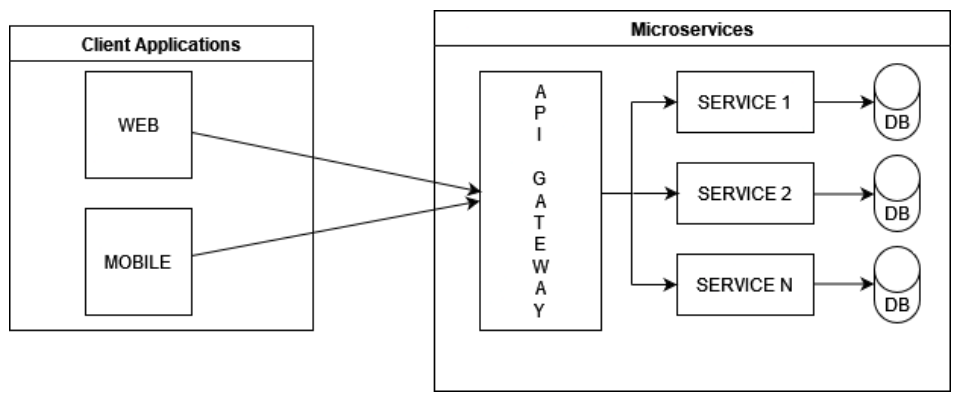

Figure 4

**Event-Driven Architecture (EDA)**: In an event-driven architecture, components communicate by generating and consuming events. Events represent significant occurrences or state changes within the system. This architecture is well-suited for asynchronous, loosely coupled systems and is commonly used in real-time data processing, IoT systems, and reactive applications.

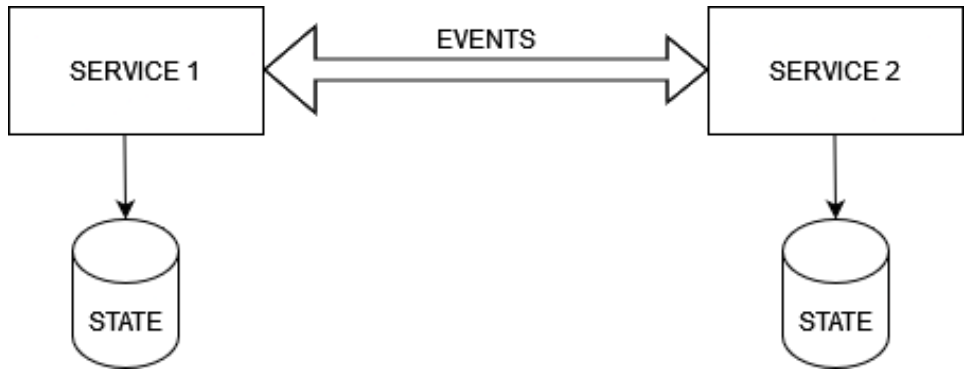

Figure 5

**Service-Oriented Architecture (SOA)**: SOA is an architectural style that structures an application as a collection of loosely coupled services. Each service provides a specific business functionality and can be accessed independently through standard protocols. SOA aims to promote reusability, interoperability, and scalability of software components.

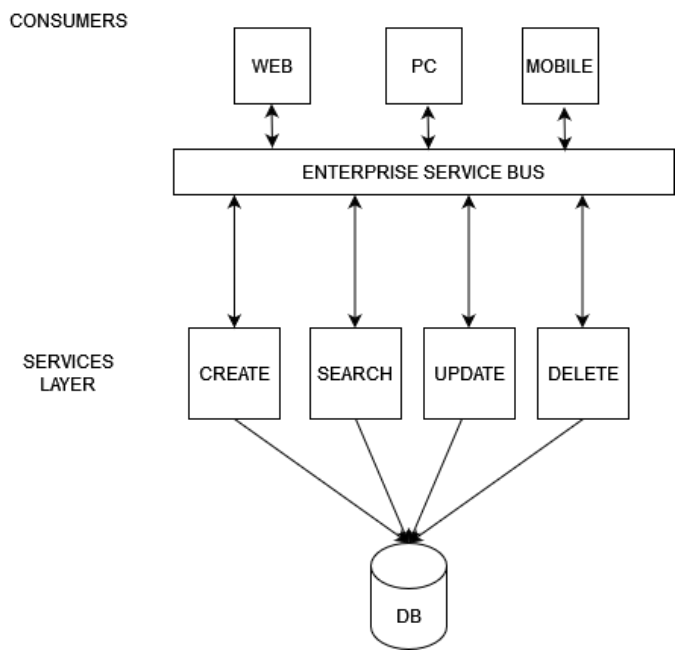

Figure 6

## 2.3 Comparison with traditional centralized systems

Centralized systems adhere to a client-server framework centered on a single server possessing substantial computational capacity. Instead of directly executing tasks, less potent nodes linked to this central server channel their process requests through it. Various applications such as Internet Service Providers, Application development servers, File Systems, and Organizational Networks exemplify centralized networks. In such setups, the distribution of computational power from the server to client machines governs system performance, contingent upon resource allocation and scheduling algorithms.

As mentioned earlier, distributed Systems operate on a peer-to-peer structure, comprising individual computers linked through a network. Each node within this system holds sufficient computational capacity to contribute to tasks collaboratively. Users within a distributed system enjoy equitable access to data, with user privileges adjustable as needed. The failure of individual components within this setup doesn't disrupt the entire system, thereby enhancing availability and reliability. Distributed systems have progressed to tackle issues encountered by traditional centralized systems, including security, data storage, and privacy concerns. Examples of large-scale distributed systems in practical use encompass the Internet, blockchain technology, and SOA-based systems.

Below is a comparison of centralized and distributed systems:

| Centralized Systems | Distributed Systems |
| --- | --- |
| Fault tolerance is absent as the central server acts as a single point of failure. | High fault tolerance as nodes can be replaced in the network without affecting performance. |
| Maintenance costs are very low as a single server needs to be monitored and manage. | Very high maintenance costs as each node needs to be monitored and managed. Moreover, the nodes could have different resources and could be distributed across geographical regions. |
| Systems can only be vertically scaled. Processing power can be added to a central server. | Systems can be horizontally and vertically scaled. Servers can be added or removed with varying loads. |
| Less throughput as the server can be a bottleneck. | High throughput because of the number of nodes participating in the system. |
| Less reliable because of single point of failure. | High reliability as system can withstand loss of nodes. |
| Complexity of system is less because only a single node is used for the system. | System is more complex because of the number of nodes participating. |

## 2.4 Applications of distributed systems in Fintech

After describing in detail and comparing distributed systems with centralized systems, let us look at different applications of distributed systems in Fintech.

### 2.4.1. Blockchain Technology in Fintech

*Overview of blockchain technology:*

Blockchain technology is a decentralized digital ledger system that records transactions across a network of computers in a secure and transparent manner. It represents a novel approach to building decentralized, trustless, and resilient distributed systems. Figure 7 illustrates an example of a blockchain. Every block references the block immediately preceding it using a hash value derived from the previous block, referred to as the parent block. The block stores data in the form of transactions, which are cryptographically secured and linked to the previous block in the chain, forming an immutable record of transaction history.

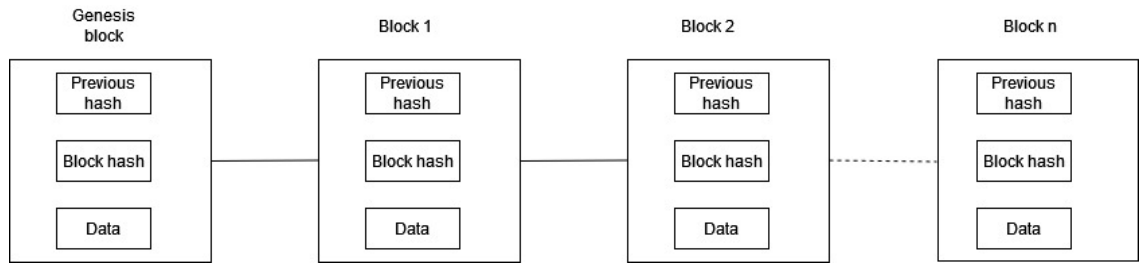

Figure 7

The block in a block chain is itself composed of two parts – header and body [6]. The block header includes:

- Version: Specifies the set of rules for validating blocks.
- Previous block hash: A 256-bit cryptographic hash value that references the preceding block.
- Merkle root: The hash value that is computed using all the transactions in the block.
- Difficulty level: Challenge required to successfully mine a new block
- Nonce: A 4-byte field, typically initialized with 0 and incremented for each hash calculation.
- Timestamp: Current timestamp

The block body comprises a transaction counter and the transactions themselves. The maximum number of transactions a block can accommodate is determined by both the block size and the size of individual transactions. Blockchain employs asymmetric cryptography to verify the authenticity of transactions. In an untrusted environment, a digital signature based on asymmetric cryptography is utilized for authentication [6].

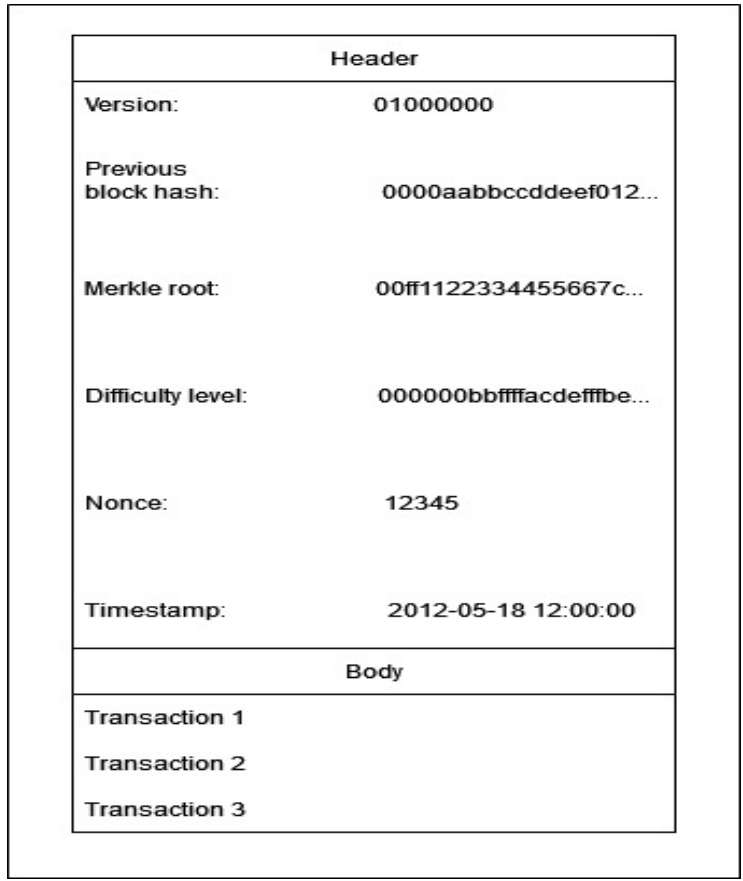

Figure 8

Blockchain functions within a decentralized framework, leveraging key technologies like digital signatures, cryptographic hashing, and distributed consensus algorithms. Transactions are processed in a decentralized manner, removing the necessity for intermediaries to validate or confirm them [7]. Blockchain possesses several fundamental traits, including decentralization, transparency, immutability, and auditability [8].

In a decentralized blockchain network, a node begins a transaction by utilizing private key cryptography to generate a digital signature. Each transaction represents the transfer of digital assets between peers on the network and is initially stored in an unconfirmed transaction pool.

These transactions are then disseminated throughout the network using a Gossip protocol, where peers must select and validate them based on predefined criteria. This validation process involves checking if the initiator possesses sufficient balance and guarding against double spending, where the same input amount is used for multiple transactions.

Once validated by miners, transactions are included in a block. Miners, who dedicate computational power to mining, solve a computational puzzle and expend computing resources to publish a block. The first miner to solve the puzzle earns the right to create a new block and receives a small incentive. Peers then verify the new block using a consensus mechanism, enabling the network to agree on its validity. The new block is added to the existing chain, confirming the transaction and creating a link to the previous block through a cryptographic hash pointer. With each subsequent block added to the chain, the transaction receives further confirmation [8].

*Applications of blockchain in FinTech:*

Blockchain technology has revolutionized the FinTech industry by providing a decentralized, transparent, and secure method for conducting financial transactions and storing data. Here are some key applications of blockchain in FinTech [9]:

i. *Cryptocurrencies*

Bitcoin, Ethereum, and other cryptocurrencies are the most well-known applications of blockchain technology. They provide a decentralized digital currency system that allows peer-to-peer transactions without the need for intermediaries such as banks. Key benefits include:

- Reduced Transaction Costs: Eliminating intermediaries reduces fees associated with transactions.
- Increased Security: Cryptocurrencies use cryptographic techniques to secure transactions.
- Global Accessibility: Cryptocurrencies can be accessed and used by anyone with an internet connection, promoting financial inclusion.

ii. *Smart Contracts*

Smart contracts are self-executing contracts with the terms of the agreement directly written into code. They automatically enforce and execute the terms when predefined conditions are met. Applications include:

- Automated Payments: Payments can be automatically released when conditions are satisfied.
- Supply Chain Management: Smart contracts can automate and verify stages in the supply chain.
- Legal Agreements: Contracts can execute clauses automatically, reducing the need for legal intermediaries.

iii. *Cross-Border Payments*

Blockchain enables more efficient and cost-effective cross-border payments by reducing the reliance on correspondent banks and intermediaries. Some of its advantages are:

- Lower Fees: Reducing the number of intermediaries lowers transaction costs.
- Faster Transactions: Blockchain transactions can settle in minutes rather than days.

- Enhanced Transparency: Transactions are recorded on a public ledger, increasing transparency.

iv. *Digital Identity Verification*

Blockchain can be used to create secure and verifiable digital identities. Applications in FinTech include:

- KYC (Know Your Customer): Blockchain can streamline KYC processes by providing a single, immutable record of a user's identity.

- Fraud Prevention: Immutable identity records help reduce identity theft and fraud.

v. *Asset Tokenization*

Blockchain enables the tokenization of physical and digital assets, allowing them to be traded more efficiently. Examples include:

- Real Estate: Properties can be tokenized, enabling fractional ownership and easier transfer of assets.

- Securities: Stocks and bonds can be issued and traded on blockchain platforms, reducing settlement times and increasing transparency.

vi. *Insurance*

Blockchain can improve the insurance industry by automating claims processing and enhancing transparency. Some of its elements include:

- Smart Contracts: Automate claims processing and payout when conditions are met.

- Fraud Detection: Immutable records help in verifying claims and preventing fraud.

- Peer-to-Peer Insurance: Decentralized platforms allow individuals to pool risk without traditional insurers.

vii. *Regulatory Compliance*

Blockchain can help financial institutions comply with regulations by providing a transparent and immutable record of transactions. Its features are:

- Audit Trails: Every transaction is recorded and cannot be altered, making it easier to audit financial activities.

- Real-Time Monitoring: Regulators can monitor transactions in real-time, improving oversight and compliance.

viii. *Payments Infrastructure*

Blockchain can enhance traditional payment systems by providing a more efficient and secure infrastructure. Key aspects include:

- Instant Settlements: Payments can settle instantly, reducing the risk and cost associated with delayed settlements.
- Lower Operational Costs: Reducing the need for multiple intermediaries lowers operational expenses.

ix. Supply Chain Finance

Blockchain can optimize supply chain finance by providing real-time visibility and transparency. Applications include:

- Invoice Financing: Real-time tracking of invoices and payments improves the efficiency of invoice financing.

- Trade Finance: Blockchain can streamline trade finance processes, reducing paperwork and fraud.

In summary, blockchain technology offers numerous benefits and applications in the FinTech industry, from enhancing payment systems and reducing costs to increasing transparency and security across various financial services [9].

**2.4.2 Decentralized Finance (DeFi)**

*Introduction to DeFi and its principles*

Decentralized finance (DeFi) is a new financial technology that utilizes secure distributed ledgers akin to those employed by cryptocurrencies. In the United States, centralized financial institutions such as banks and brokerages operate under regulations set by the Federal Reserve and the Securities and Exchange Commission (SEC), providing consumers with direct access to capital and financial services. DeFi disrupts this traditional centralized system by enabling individuals to conduct peer-to-peer transactions [10]. DeFi is not an official legal or technical term. However, it is increasingly prevalent in discussions about the future of finance and its regulation. Typically, DeFi involves aspects such as: (i) decentralization; (ii) distributed ledger technology and blockchain; (iii) smart contracts; (iv) disintermediation; and (v) open banking [11].

DeFi arises from three key trends in technological evolution: Moore's law, Kryder's law, and a third, yet unnamed pattern. Moore's law suggests that data processing power increases exponentially, while Kryder's law asserts the same for data storage capacity. These advancements lead to continuously decreasing costs for processing and storage. The third trend involves the rapid growth of communications bandwidth paired with decreasing costs, a concept discussed since the late 1990s. This growth is driven by enhanced network efficiencies, resulting in more bandwidth per dollar through innovations like lower production costs of network components, denser and faster ports, higher utilization, integrated photonics, and higher frequency microwaves with smaller cells using multiple frequency bands (e.g., 5G).

These three trends enable hardware virtualization, where software is hosted, updated, and run on decentralized servers instead of individual workstations. Local data processing only occurs when necessary, thanks to constant online connectivity and abundant bandwidth. Hardware virtualization facilitates the development of service-oriented architecture ('software as a service'), which is central to DeFi. Concurrently, Moore's law, Kryder's law, and decreasing bandwidth costs continue to drive progress, supporting advancements in machine learning, AI, and 'edge' computing. This includes substantial localized processing in devices like those in the Internet of Things (IoT), which utilize both virtual and local data and processing power. These technological trends and the intellectual processes enabling their integration are transforming finance and numerous other fields [11].

At the heart of DeFi are several emerging technologies collectively referred to by the acronym 'ABCD,' which encapsulates the core of FinTech. These technologies are AI, Blockchain (and distributed ledgers and smart contracts), Cloud, and Data (both big and small). Alternatively, ABCD can also stand for AI, Big Data, Cloud, and DLT (including blockchain and smart contracts) [11].

These four rapidly evolving technologies are central to the decentralization of finance due to their specific applications in this area. Decentralized financial functions leverage: (i) the efficiencies and cost savings of AI; (ii) the advanced record-keeping and efficiencies of smart contracts on distributed ledgers secured by blockchain; (iii) the powerful capabilities of algorithmic data analysis; and (iv) cloud systems to host virtually all decentralized financial functions. Each of these technologies benefits from the previously mentioned technological 'laws,' as they become increasingly cost-effective, convenient, and efficient. This enables cooperation among multiple participants who collectively provide decentralized financial services [11].

*Applications of DeFi*

The principles discussed above have been implemented in a number of applications. In this section, we provide a brief overview of those applications.

i) *Decentralized Exchanges*: Facilitating decentralized asset exchange requires an efficient method for matching buyers and sellers at specific prices, known as price discovery. Early decentralized exchanges (DEXs) on permissionless blockchains demonstrated that executing decentralized asset exchanges using a central limit order book (CLOB) design was feasible. However, this approach proved impractical and costly at scale for several reasons [12].

First, in the blockchain-based virtual machine format, traders pay fees based on the complexity and storage needs of their transactions. Since the virtual machine replicates across all nodes, storing even small amounts of data becomes very expensive. The complex matching logic needed to maintain a liquid order book further increases computational fees, surpassing users' willingness to trade [12].

Second, miners select transactions for the next block based on attached computational fees. This allows for front-running state changes in the decentralized order book by attaching high fees to transactions, exploiting the next state change for arbitrage opportunities [12].

To address these issues, subsequent DEX iterations stored the order book state separately and used the blockchain solely for final settlement. However, this introduced coordination problems among order book storage providers, increasing security risks [12].

In response to the limitations of the CLOB design, a new generation of blockchain-specific 'automated market makers' (AMMs) emerged. AMMs pool liquidity in trading pairs or groups, eliminating the need for simultaneous buyer and seller presence and enabling seamless trade execution without compromising blockchain integrity. Liquidity providers supply crypto assets in exchange for trading fee returns [12].

ii) *Peer-to-Peer Lending and Algorithmic Money Markets:* Money markets, which involve borrowing and lending capital with interest payments, play a crucial role in traditional finance. Within DeFi, borrowing and lending applications are among the largest financial segments, with $7 billion in total value locked by the end of 2020. In these protocols, agents with excess capital, known as liquidity providers, lend crypto assets to a peer-to-peer protocol and receive continuous interest payments. Borrowers, in turn, borrow crypto assets and pay interest [12].

Due to the pseudonymous nature of blockchain, borrowing purely on credit is not possible. Borrowers must overcollateralize their loans by providing collateral in crypto assets that exceed the loan's value. The smart contract then issues a loan amounting to 70-90% of the collateral's value. If the collateral value drops below the outstanding loan's value, the smart contract automatically auctions the collateral on a decentralized exchange at a profit. Interest rates are set algorithmically based on the supply and demand for each specific crypto asset [12].

Initially introduced by the MakerDAO application, several protocols now offer similar services with novel interest rate calculations or optional insurance features, currently managing $7 billion in crypto assets [12].

iii) *Derivatives:* Blockchain-based financial contracts, particularly derivatives, are one of the fastest-growing segments in DeFi. Developers aim to integrate traditional financial derivatives, such as options, futures, and other synthetic contracts, into the broader DeFi ecosystem. A futures contract involves agreeing to sell an asset at a specific price on a future date, while an options contract provides the right, but not the obligation, to buy or sell an asset at a set price [12].

Similar to traditional finance, these services can be used for market movement insurance or price speculation. Recently, a new category of 'synthetic' assets has emerged, represented by tokens pegged to external prices, commonly tracking commodities (like gold) or stocks (like Tesla). Users can create synthetic assets by collateralizing crypto assets in a smart contract, similar to decentralized lending. These synthetic assets follow an external price feed, known as an 'oracle,' provided to the blockchain. However, these price feeds can face technical issues and coordination problems, leading to delays or manipulation, especially during network congestion [12].

iv. *Automated Asset Management:* In traditional financial services, asset management involves allocating financial assets to meet the long-term financial goals of individuals or institutions. In DeFi, numerous applications now perform this function algorithmically, without human intervention, allowing markets to operate continuously and beyond manual control [12].

Automated asset managers in DeFi primarily serve two purposes: 'yield aggregators' and crypto asset indices. Yield aggregators are smart contract protocols that allocate crypto assets based on predefined rules to maximize yield while managing risk. Users deposit assets into these protocols, which then distribute the assets across various applications to optimize returns and continuously rebalance allocations [12].

Crypto asset indices, similar to passive investing, provide broad exposure to a portfolio of crypto assets. These indices automatically purchase and hold a selection of assets within a smart contract. Stakeholders can buy a token representing ownership of the index, similar to exchange-traded funds (ETFs), giving them algorithmic rights over a portion of the total assets held [12].

*Advantages and risks associated with DeFi applications*

Decentralized finance (DeFi) is a new financial model that uses blockchain technology and smart contracts to deliver financial services without the need for intermediaries. Some of its advantages are:

i. *Decentralization:* DeFi operates on decentralized networks, particularly blockchain, a distributed ledger technology that securely, transparently, and immutably records transactions across multiple computers, as well as peer-to-peer networks. These technologies function without central authorities or intermediaries [13].

ii. *Smart Contracts:* DeFi depends on smart contracts, which are self-executing contracts with terms directly written into code running on a blockchain. These contracts are tamper-proof and automatically enforceable without needing a centralized authority. They enable DeFi to automate complex operations, eliminate intermediaries, and enhance efficiency and trust [13].

iii. *Crypto Assets:* DeFi utilizes crypto assets, such as cryptocurrencies and tokens, for exchanging and storing wealth. These assets provide DeFi products with greater liquidity, flexibility, speed, and accessibility compared to traditional assets, allowing for innovative financial products and services [13].

iv. *Open Finance:* Open finance involves using open protocols, standards, and networks to facilitate financial innovation and interoperability. This enables DeFi to offer more inclusive, interoperable, and decentralized financial services than traditional finance [13].

After comparing its advantages, let us compare its risks:

i. *Smart Contracts:* Developing smart contracts involves translating code into programming languages, which can introduce exploitable vulnerabilities and errors, including arithmetic vulnerabilities, re-entrancy, block randomness issues, and overcharging. Ensuring smart contracts are bug-free before deployment is crucial, but verifying their correctness is challenging due to their complexity. This verification is vital to avoid exploitation. Smart contract execution depends on real-world data from trustworthy oracles. Transaction order dependencies and execution serialization can limit performance. Some systems, like software transactional memory, aim to improve efficiency [13].

ii. *Completion Risks:* Privacy concerns arise as most blockchain transactions are publicly visible despite pseudonymous public keys. This visibility lacks privacy-preserving mechanisms, leading to potential data exposure [13].

iii. *Risks from the Blockchain Protocol:* Blockchain protocols can change via hard forks, soft forks, or governance processes, posing risks of fraud or sub-optimal outcomes. Consensus protocols like PoW or PoS may lead to centralization, where a few users control the network's computational power, risking 51% attacks and other issues like selfish mining and Sybil attacks. DeFi products may suffer from delays and inefficiencies if the blockchain protocol cannot handle high transaction volumes or has high latency [13].

iv. *Front Running:* Front running involves exploiting knowledge of future transactions for profit, using methods like sandwich attacks in the transaction pool (mempool) [13].

v. *Oracle Vulnerabilities:* Oracles provide smart contracts with external data, making them susceptible to manipulation through technical or social vulnerabilities. Examples include padding oracle attacks and price data manipulation, as seen in the Solend lending platform hack [13].

vi. *Data Authenticity Solutions:* Solutions like PADVA and TownCrier aim to provide cryptographically verifiable data, but they have limitations, such as centralized points of failure and the potential for websites to manipulate outputs [13].

vii. *Impermanent Loss:* During market fluctuations, liquidity pools may face impermanent loss, where the fiat value of deposited crypto assets decreases over time [13].

viii. *Smart Contract Vulnerabilities:* Once assets are in a liquidity pool, they are controlled by a smart contract. Exploiting bugs or vulnerabilities can result in permanent loss of assets [13].

ix. *Liquidation Risk:* Leveraged liquidity pools risk forced liquidation if asset prices drop. Recent incidents, like the near-liquidation of Solend's SOL deposits, highlight this risk [13].

x. *Flash Loan Attacks:* Flash loans, non-collateralized loans repaid in the same transaction, can be used for market manipulation, leading to significant asset theft [13].

xi. *Vampire Attacks:* These attacks drain liquidity from one exchange to another. An example is SushiSwap siphoning $1.2 billion in liquidity from Uniswap [13].

xii. *Cybersecurity Threats:* Hackers can steal assets by accessing users' private keys or exploiting smart contract flaws. DDoS attacks can render DeFi platforms inaccessible, disrupting transactions [13].

**2.4.3 Distributed Ledger Technology (DLT)**

*Overview of DLT*

Distributed ledger technology (DLT) encompasses the set of infrastructure and protocols enabling concurrent access, validation, and updating of records within a networked database. It serves as the foundational framework for blockchains and enables users to track alterations and their sources, thereby minimizing the necessity for extensive data auditing. DLT ensures the integrity of data and restricts access solely to authorized parties [14].

DLTs aim to facilitate interactions among users who lack mutual trust, eliminating the necessity for a trusted intermediary. This is particularly crucial in scenarios where there's inherent distrust among participants, such as between business partners or anonymous entities. DLTs inherently provide transparency, traceability, and security in such environments. Essentially, DLTs serve as data structures for recording transactions and offer functions for managing them. While each DLT may employ distinct data models and technologies, they generally rely on three foundational technologies: public key cryptography, distributed peer-to-peer networks, and consensus mechanisms, which are combined uniquely. Public key cryptography is utilized to establish secure digital identities for participants, with each participant possessing a pair of keys (public and private) for transaction recording in the Distributed Ledger (DL). These digital identities ensure ownership control over managed objects within the DL. Employing a peer-to-peer network enables scalability, mitigates single points of failure, and prevents dominance by a single entity or small group. A consensus protocol enables all DL participants (nodes) to converge on a single truth without relying on a trusted third party. Some distributed ledger technologies are Blockchains, Tangle, Hashgraph & Sidechains. [15]. As blockchain has already been described, this paper will give an overview of Tangle, Hashgraph & Sidechain.

*Tangle:*

The Tangle, developed by IOTA, is a blockchain alternative specifically designed for the Internet of Things (IoT). It uses a Directed Acyclic Graph (DAG) data structure, where each node represents a transaction, and connections between nodes indicate transaction validations. Unlike traditional blockchains, Tangle has no transaction fees and merges the roles of transaction makers and validators, allowing all users to validate transactions, which is crucial for micro-transactions in IoT.

Key features of Tangle include:

- Scalability: Validation time improves as more users join, and the required computational power is low, making it accessible for devices like smartphones and IoT devices.
- Quantum-Resistance: Tangle employs the Winternitz One-Time signature scheme, which is resistant to quantum computing attacks, enhancing security.
- No Transaction Fees: By eliminating the distinction between transaction makers and validators, users use their computing power to validate transactions without paying fees.

In Tangle, users must validate two previous transactions using a Proof of Work (PoW) computation, which helps prevent spam transactions. This system encourages network participation and ensures secure, scalable, and fee-free transactions [15].

*Hashgraph:*

Hashgraph is seen as an evolution of blockchain, addressing its limitations in speed, fairness, cost, and security. Developed by Mance Harmon, Hashgraph employs a Directed Acyclic Graph (DAG) for transaction storage and a combination of a voting algorithm and gossip protocol to achieve rapid consensus.

Key Features:

- Consensus Protocol: Hashgraph's consensus protocol is mathematically proven to replicate data much faster than blockchain.
- Structure: The network consists of users (columns) and events (vertices). Users can submit transactions and gossip about them to others, enabling exponential information spread.
- Information Handling: Each event includes hashes of previous events, transactions, and timestamps, with a voting algorithm to detect any timestamp forgery.
- Fairness and Ordering: Ensures events are processed in the order they are submitted, addressing resource access conflicts more fairly than blockchain.
- Virtual Voting: Every node holds a full copy of the hashgraph and can calculate other nodes' votes without transmitting them, ensuring efficiency even when not all nodes are synchronized.
- Efficiency: No events become stale or discarded, and the gossip protocol reduces communication overhead, supporting a high transaction rate. However, Hashgraph has primarily been tested in private settings.

Hashgraph's unique contributions include a fair network with timestamped consensus, virtual voting, and efficient handling of transactions, positioning it as a significant advancement in Distributed Ledger Technology (DLT) [15].

*Sidechain:*

Sidechains enhance blockchain technology by combining two blockchain architectures to address limitations in security, privacy, and performance. This setup uses a central consortium blockchain for managing access requests and multiple private sidechains for local transactions. Each local chain can seclude its information and selectively share data, improving privacy and control.

Key Features and Benefits:

- Improved Scalability: Sidechains divide the network into sub-networks, where each sub-network validates only its transactions, reducing the load on the overall network.
- Privacy: Sidechains allow private transactions within a group of partners while hiding data from competitors. Validators in each sidechain connect with the central consortium to manage local transactions and access logs.

- Customizable Access Control: The architecture allows for private channels with strict control over data access, enhancing privacy while maintaining blockchain's inherent security features.

Sidechains thus offer a flexible and scalable solution for blockchain networks, balancing local privacy needs with broader network connectivity and security.

## 3. Future Directions

Financial Technology (FinTech) has significantly evolved due to advances in distributed systems. Innovations in blockchain, distributed ledger technologies (DLTs), smart contracts, and consensus mechanisms have paved the way for more efficient, secure, and transparent financial services.

With these advancements we have a unique set of challenges and opportunities that lie ahead of us. We have seen that Blockchain's immutable ledger ensures that all transactions are securely recorded and cannot be altered. This enhances the security and transparency of financial transactions. Moreover, as all the transactions are public this ensures that there is no fraud taking place in the system. As we have seen previously, distributed technologies can significantly reduce the time required for processing transactions by eliminating the need for intermediaries. This characteristic paves the way for faster transactions and lower costs. By automating and streamlining processes, financial institutions can reduce operational costs. Another compelling feature of distributed technologies more specifically decentralized finance (DeFi) is that DeFi platforms can offer financial services like lending, borrowing, and trading without traditional banks, potentially increasing access and reducing barriers. This can bring financial inclusion to underserved populations, especially in regions with limited access to traditional banking infrastructure. Similarly, assets like real estate, stocks, and art can be tokenized, making them more liquid and accessible to a broader range of investors. Finally, distributed ledgers provide real-time access to financial records, facilitating more efficient auditing processes. Another key innovation in distributed technologies is that of smart contracts. The advantage of smart contracts is that it can automatically enforce regulatory compliance, reducing the burden on financial institutions and enable programmable money that can execute complex financial transactions automatically.

## 4. Conclusion

Distributed systems, particularly blockchain and distributed ledger technology (DLT), are set to transform the FinTech industry by decentralizing financial services, enhancing security, and driving innovation. These technologies enable peer-to-peer financial services, improve financial inclusion by providing access to underserved populations, and ensure the integrity of transactions through immutable ledgers and smart contracts. The increased transparency and real-time auditing capabilities of distributed systems can simplify regulatory compliance, reduce fraud, and enhance trust among users. Additionally, the ability to tokenize assets and implement programmable money opens up new avenues for financial products and services.

However, significant challenges must be addressed for widespread adoption. Scalability issues, performance limitations, and high energy consumption of current systems pose technical hurdles. Regulatory uncertainty and the need for interoperability with existing financial infrastructure also present obstacles. To navigate these challenges, FinTech companies should invest in research and development, engage proactively with regulators, and collaborate with traditional financial institutions and technology providers. By focusing on these strategic areas, the FinTech industry can leverage distributed technologies to create a more efficient, secure, and inclusive financial ecosystem.

In conclusion, distributed systems are pivotal in advancing the security, scalability, and efficiency of financial operations. Their continued development and integration hold the promise of a more robust, agile, and inclusive financial ecosystem.

# References

[1] Gomber, P., Koch, J, A and Siering, M Digital Finance and Fintech: current research and future research directions, J Bus Econ, Vol. 87, p.p. 537-580, 2017

[2] Sarhan, H. Fintech: An Overview. Available online: https://www.researchgate.net/publication/342832269

[3] van Steen, M., Tanenbaum, A.S. A brief introduction to distributed systems. Computing 98, 967–1009, 2016

[4] Jeffrey Joyce, Greg Lomow, Konrad Slind, and Brian Unger. 1987. Monitoring distributed systems. ACM Trans. Comput. Syst. 5, 2, pp. 121–150, 1987.

[5] Le Lann, Gérard. "Distributed Systems-Towards a Formal Approach." In *IFIP congress*, vol. 7, pp. 155-160, 1977.

[6] Z. Zheng, S. Xie, H. N. Dai, X. Chen, and H. Wang, "Blockchain challenges and opportunities: a survey," *International Journal of Web and Grid Services*, vol. 14, no. 4, pp. 352–375, 2018, doi: https://doi.org/10.1504/ijwgs.2018.095647.

[7] S. Nakamoto et al., Bitcoin: A Peer-to-Peer Electronic Cash System. Citeseer, 2008. [Online]. Available: http://bitcoin.org/bitcoin.pdf

[8] Monrat, Ahmed Afif, et al. "A Survey of Blockchain from the Perspectives of Applications, Challenges, and Opportunities." IEEE Access, vol. 7, no. 7, 2019, pp. 117134–117151, ieeexplore.ieee.org/abstract/document/8805074, https://doi.org/10.1109/access.2019.2936094.

[9] Renduchintala, Tara, et al. "A Survey of Blockchain Applications in the FinTech Sector." Journal of Open Innovation: Technology, Market, and Complexity, vol. 8, no. 4, 13 Oct. 2022, p. 185, https://doi.org/10.3390/joitmc8040185.

[10] R. Sharma, "Decentralized finance (Defi) Definition and Use Cases," *Investopedia*, Dec. 05, 2023. https://www.investopedia.com/decentralized-finance-defi-5113835

[11] D. A. Zetzsche, D. W. Arner, and R. P. Buckley, "Decentralized Finance (DeFi)," SSRN Electronic Journal, vol. 6, no. 2, 2020, doi: https://doi.org/10.2139/ssrn.3539194.

[12] J. R. Jensen, V. von Wachter, and O. Ross, "An Introduction to Decentralized Finance (DeFi)," Complex Systems Informatics and Modeling Quarterly, no. 26, pp. 46–54, Apr. 2021, doi: https://doi.org/10.7250/csimq.2021-26.03.

[13] T. Weingärtner, F. Fasser, P. Celestino, and W. Farkas, "Deciphering DeFi: A Comprehensive Analysis and Visualization of Risks in Decentralized Finance," Journal of risk and financial management, vol. 16, no. 10, pp. 454–454, Oct. 2023, doi: https://doi.org/10.3390/jrfm16100454.

[14] J. Frankenfield, "Distributed Ledger Technology," Investopedia, Aug. 27, 2021. https://www.investopedia.com/terms/d/distributed-ledger-technology-dlt.asp

[15] N. El Ioini and C. Pahl, "A Review of Distributed Ledger Technologies," Lecture Notes in Computer Science, pp. 277–288, 2018, doi: https://doi.org/10.1007/978-3-030-02671-4_16.